\documentclass[11pt,reprint,a4paper]{article}
\usepackage{eurosym}
\usepackage[affil-it]{authblk}
\usepackage{setspace}
\usepackage{color}
\usepackage[dvipsnames]{xcolor}
\usepackage{amsmath}
\usepackage{amsfonts}
\usepackage[title]{appendix}
\usepackage{comment}
\usepackage{booktabs}
\usepackage{verbatim}
\usepackage{amssymb}
\usepackage{float}
\usepackage{subcaption}
\usepackage{physics}
\usepackage{mathrsfs}
\usepackage{comment}
\usepackage[colorlinks]{hyperref}
\hypersetup{
    colorlinks=true,
    linkcolor=blue,
    filecolor=magenta,
    urlcolor=[rgb]{0.00,0.00,0.50},
    citecolor=red,
    }
\usepackage[top=2cm, bottom=2cm, left=3cm, right=3cm]{geometry}

\usepackage[normalem]{ulem}

\usepackage[backend=bibtex,sorting=none,style=trad-abbrv,doi=false,url=true]{biblatex}
\DeclareMathAlphabet\mathbfcal{OMS}{cmsy}{b}{n}
\DeclareMathAlphabet{\boldmathe}{T1}{cmr}{bx}{it}

\def\be{\begin{equation}}
\def\ee{\end{equation}}

\def\N{\mathbb Z}
\def\N{\mathbb N}
\def\R{\mathbb R}

\def\be{\begin{equation}}
\def\ee{\end{equation}}

\def\N{\mathbb Z}
\def\N{\mathbb N}
\def\R{\mathbb R}

\def\be{\begin{equation}}
\def\ee{\end{equation}}

\def\N{\mathbb N}
\def\R{\mathbb R}

\def\Ti{\text{i}}

\usepackage{todonotes}

\definecolor{darkgreen}{rgb}{0,0.70,0}

\newcommand{\USACh}{Departamento de F\'isica, Universidad de Santiago de Chile,  Av. Victor Jara 3493, Santiago, Chile}
\newcommand{\USAL}{IUFFyM, Universidad de Salamanca, Salamanca E-37008, Spain}
\newcommand{\UACh}{Instituto de Ciencias F\'isicas y Matem\'aticas, Universidad Austral de Chile, Casilla 567, 5090000 Valdivia, Chile}
\newcommand{\UCharles}{IPNP - Faculty of Mathematics and Physics, Charles University, V Hole\v{s}ovi\v{c}k\'ach 2, 18000 Prague 8, Czech Republic}

\begin{document}

\title{\bf Confining kinks. $\zeta$-regularized \\ one-loop kink mass shifts in exotic field theories}
\author[1]{Luis Inzunza\thanks{luis.inzunza@usach.cl}}
\author[2]{Juan Mateos Guilarte\thanks{guilarte@usal.es}}
\author[3,4]{Pablo Pais\thanks{pais@ipnp.troja.mff.cuni.cz}}

\affil[1]{\USACh} \affil[2]{\USAL} \affil[3]{\UACh} \affil[4]{\UCharles}

\date{}
\maketitle

\begin{abstract}
By combining stability and the analysis of  topological solitons in $(1+1)$-dimensional scalar field theories with the Darboux transformation technique, we create models featuring kink-like solutions whose perturbations are all bounded. The stability analysis of extended classical solutions in scalar field theories is determined from the spectrum of quantum mechanical Schr\"odinger operators whose eigenfunctions serve to expand quantum perturbations to the classical soliton. Our strategy is to invert the procedure: we start from the Schrödinger operator for a single harmonic oscillator and ask whether there is a field theory admitting kink solutions. Having identified the parent exotic field theory to the harmonic oscillator, the Darboux transformation allows for constructing new exotic but solvable Schr\"odinger equations. This framework relates the quantum harmonic oscillator and its rational deformations to exotic scalar theories featuring non-trivial potentials. Depending on the structure of the spectrum of perturbation frequencies, these potentials may have various local maxima, minima, and inflexion points. The stationary solutions take the form of a definite integral over a finite interval of a function times the Gaussian bell distribution, including the error and the Owen $T$ functions. In such models, the spectrum of the Schr\"odinger operator governing kink fluctuations is purely discrete. Therefore, Riemann$-\zeta$ function regularization allows us to achieve finite one-loop quantum corrections to the classical mass.
\end{abstract}

\newpage

\section{Introduction}

Solitons are non-linear waves that maintain their shape after scattering and suffer only mild phase shifts. Even the weaker concept of solitary waves that change their shape after collisions occupies a very special place in field theory, condensed matter, fluid dynamics, and cosmology, among others, as extended objects even at the classical level, while descending to extended quantum lumps. To mention one case, in the Skyrme model Skyrmions play the role of baryons, whereas fundamental quanta describe mesons \cite{Skyrme}.

Because solitons are classical solutions in field theory, their stability is tested by their response to linear perturbations. It is well known that linear perturbations to classical solutions encode quantum corrections if classical solutions are stable.  Thus, one expects that stability analysis brings with it some renormalization of the classical parameters at the classical level, for example, classical mass. Therefore, studying such perturbations is paramount, and many approaches exist to obtain these corrections.

Even if they are still not considered fundamental theories, solitonic solutions in toy models of field theory are of considerable interest, as they can provide mathematical insight, either analytical or numerical, about complicated phenomena otherwise difficult to deal with, such as confinement or quantum quench in integrable and non-integrable models (see, for instance, \cite{Kormos2016}).

Among the solitons, kinks play an important role, as they appear in the simplest models in $(1+1)$-dimensions (see \cite{Rubakov}, and references therein). As topological solitons, kinks have been extensively used as toy models for complicated phenomena. As a stable extended object, studying perturbations around the classical solution of kinks has recently received considerable attention some time ago \cite{Osman1995}, with some renewed interest recently \cite{Hartmann2020,sineGordonMaps2023}. The different field theoretical models are characterized by the part of the energy independent of the field derivatives. These polynomial or transcendental functions are chosen in such a way that the static field equations support kink solutions. In this paper, we study the inverse problem, i.e., we start from Schr$\ddot{\rm o}$dinger operators with harmonic oscillator and/or its rational extensions
as potentials and search for the field theoretical parts of the energy density that correspond to these kink fluctuation analysis.
By construction, we shall deal with models where all the kink fluctuations are bounded.

To address this issue, we start with the key observation that perturbations of $(1+1)$ dimensional scalar theories obey Schrödinger-like equations \cite{Rubakov}. At this point, the Darboux transformation enters the construction naturally \cite{Cooper, Matveev}. This technique enables the generation of new equations and their solutions from known equations and solutions. We can use this method to construct new families of exactly solvable Schrödinger equations from some well-known ones, and then identify their solutions as perturbations of new and distinct scalar theories. Now, the transformation is characterized by a non-trivial kernel and therefore is not a one-to-one mapping. Consequently, the new Schrödinger operator is almost isospectral to the original one, but it can exhibit a specific fine structure, such as isolated bound states or forbidden bands. These features reveal themselves in the scalar final form of the model and are crucial when considering the quantization of the theory. Bounded perturbations corresponding to the discrete part of the spectrum are linked with confined particles trapped by the kink, while scattering perturbations would be associated with propagating mesons. The Dashen-Hasslacher-Neveu (DHN) formula \cite{dhn1,dhn2} for computing the one-loop mass correction of the kink encompass bound and scattering waves fluctuations. In our models there are only bounded perturbations to the kink. Thus we are led to deal with a simplified version of the DHN formula.

In this work, we focus on the family of rational extended harmonic oscillator potentials with gaps (holes) in the equidistant discrete spectrum \cite{quesne,quesne2, Gomez, Carinena:2016hfq, Carinena:2017bqs, Carinena:2017zfy} as the Schrödinger operator that controls the perturbations' behaviour. As a result, we obtain scalar models characterized by kinks with pure integral representation (e.g., the error and the Owen $T$ functions), which we call \emph{confining kink}, since all perturbations are bounded in the space. 
Remarkably, the space-time independent classical solutions in our models do not admit linear perturbations, neither bounded nor extended. Therefore, the DHN formula reduces only to an infinite discrete sum. Regarding the regularization methods, one available technique is the $\zeta$-function regularization. It is a valuable tool for computing the zero-point energy manifestation in quantum field theories, such as the Casimir effect (see \cite{Bordag2001} and references therein for an excellent review). This method can also be used to compute one-loop kink mass, either in general \cite{AM1, AM2}, or for supersymmetric quantum field theories \cite{AMP}. When this technique is used for computing the one-loop mass for our confining kink, a finite negative value is obtained without the need for a renormalization scheme. Additionally, this result implies the presence of an attractive \emph{``quantum force" that reinforces the boundedness of meson} propagation in our exotic kink backgrounds.

The paper is organized as follows. In Section \ref{sec_construction}, we introduce a review about the construction of scalar fields theories containing kinks, including also one-loop corrections. Then the formalism is employed to build the first example of confining kink corresponding to the error kink model. In Section \ref{sec_new_species},  the Darboux transformation method is used to generate Schr\"odinger Hamiltonian operators with a discrete spectra bounded from below. In the same paragraph, we present models that exhibit localized fluctuations around the kink solitary waves, as the Owen kinks and deformed confining kinks, where $\zeta$-function regularization is used to shown there is no zero-point mass renormalization for such kinks. Finally, Section \ref{Sec_Discussion} is devoted to the discussion and concluding remarks.

\section{Constructing scalar field theories using stability criteria}
\label{sec_construction}
This section presents the theoretical framework for constructing scalar field models featuring confining kinks. This procedure is based on the stability analysis of the corresponding stationary solutions of a generic scalar theory \cite{AM1,TrulFle}. It also involves solving eigenvalue problems of the Schr\"odinger type, where the spectrum is positive, discrete, and bounded from below. Additionally, we discuss key aspects of one-loop quantum corrections to the classical mass of these solutions, which arise from canonical quantization \cite{AM1, AM2, AMP}.
Finally, we employ this machinery to construct the explicit error kink model together with all its classical and quantum features.

\subsection{Classical picture}
Before discussing what a confining kink is and how to build it, it is worth considering the action for a general $(1+1)$ scalar field theory ($c=1$)
\begin{equation}
\label{action}
S[\Phi(t,x)]=\int_{\R^{1,1}} dz\,dt \left[\frac{1}{2}\left(\frac{\partial \Phi}{\partial t}\frac{\partial \Phi}{\partial t}-\frac{\partial \Phi}{\partial z}\frac{\partial \Phi}{\partial z}\right)-\mathcal{V}([\Phi(t,z)])\right]\;,
\end{equation}
where $\mathcal{V}([\Phi(t,z)])$ is the Lorentz invariant part of the potential energy density. To simplify the analysis of such system, it is worth to introduce by hand two constant parameters $\lambda$ and $\nu$ with dimensions
\begin{equation}
[\lambda]=L^{-1} \;,\qquad [\nu]=M^\frac{1}{2}L^{-1/2}\;,
\end{equation}
in terms of which we can define dimensionless variables
\begin{equation}
\phi(\tau,x)=\frac{\lambda}{\nu}\Phi(t,z)\;,\qquad V[\phi(\tau,x)]=\frac{1}{\nu^2} \mathcal{V}[\Phi(t,z)]\;,\qquad \tau=\lambda t\;,\qquad x=\lambda z\;.
\end{equation}
In consequence, the action now reads as
\begin{equation}
S[\Phi(t,z)]=\frac{\nu^2}{\lambda^2} \int_{\R^{1,1}} dx d\tau \left[\frac{1}{2}\left(\frac{\partial \phi}{\partial \tau}\frac{\partial \phi}{\partial \tau}-\frac{\partial \phi}{\partial x}\frac{\partial \phi}{\partial x}\right)-V[\phi(\tau,x)]\right]\;.
\end{equation}
The system is invariant under time translations, and the classical Hamiltonian can be obtained by means of the Legendre transformation. As a result we get
\begin{equation}
\label{Ham}
\mathcal{H}[\Phi(t,z)]=\frac{\nu^2}{\lambda} \int dx  \left[\frac{1}{2}\left(\frac{\partial \phi}{\partial \tau}\frac{\partial \phi}{\partial \tau}+\frac{\partial \phi}{\partial x}\frac{\partial \phi}{\partial x}\right)+V[\phi(\tau,x)]\right]\;.
\end{equation}
The variation of the action produces the equation of motion of the system
\begin{equation}
\label{em}
\left(-\frac{\partial^2}{\partial \tau^2}+\frac{\partial^2}{\partial x^2}\right)\phi(\tau,z)=\frac{\partial V(\phi)}{\partial \phi}\;,
\end{equation}
and for searching stationary solutions $\phi_K(\tau,x)=\varphi(x)$, we can multiply from both side of this last equation by $\frac{d}{dx}\varphi(x)$ and integrate it with respect to $x$. As a result we find, up to a constant set to zero to keep the kink energy finite, that
\begin{equation}
\label{stateq}
\frac{1}{2}\left(\frac{d \varphi}{dx}\right)^2=V(\varphi)\;.
\end{equation}
The energy associated with this solution corresponds to its mass at rest, and, with the help of expression (\ref{stateq}),
we can compute this quantity in terms of $V(\phi(x))$ as a function of $x$,
\begin{equation}
\label{MMM}
 M=\mathcal{H}[\varphi(x)]=\frac{2\nu^2}{\lambda} \int_{-\infty}^{\infty}V(x) dx \,\;,\qquad
V(x)=V[\varphi(x)].
\end{equation}
An alternative formula for this quantity in terms of the field can be obtained by performing the change of variable $x\rightarrow \varphi(x)$. The final expression is given in terms of the  so-called \emph{``superpotential"} $W=\int \, d\varphi \, \sqrt{2 V(\varphi)}$ , and reads as
\begin{equation}
\label{MW}
M= \frac{\nu^2}{\lambda} \int_{\varphi(-\infty)}^{\varphi(\infty)}\, d\varphi \, \sqrt{2 V(\varphi)} = \frac{\nu^2}{\lambda}\Big\vert W(\varphi(\infty))- W(\varphi(-\infty))\Big\vert \;.
\end{equation}
This last formula is very useful since allows to compute the mass by only fixing boundary condition, without solving equation (\ref{stateq}), and can be employed to introduce fermions (see Section \ref{Sec_Discussion}).

To study the  linear stability of the stationary  solution in the general case, we plug the small fluctuation expansion over it up to first order
\begin{align}
\label{pertur}
&\phi(\tau,x)=\varphi(x)+\alpha \, \eta(\tau,x)\,+ {\cal O}(\alpha^2),\qquad \alpha << 1\;,
\\
&\eta(\tau,x)=\sum_{n}\frac{1}{\sqrt{2\omega_n }}(a_ne^{\Ti   \omega_n \tau}+a_n^*e^{-\Ti   \omega_n \tau}) \psi_n(x)\;,
\end{align}
in the full equation of motion (\ref{em}). Here, $\omega_n$ represents a distribution of dimensionless frequencies, characterized by the index $n$, which can be either discrete or continuous. Additionally,  $\psi_n(x)$ are the so-called perturbation modes that satisfy the Schr\"odinger-like equation
\begin{equation}
\label{Sch}
\left(-\frac{d^2}{dx^2}+U(x)\right)\psi_n=\omega_n^2 \psi_n(x)\;,
\qquad
U(x)=\frac{\partial^2V(\phi)}{\partial \phi^2}\big|_{\phi=\varphi(x)}\;.
\end{equation}
As long as the dimensionless frequencies $\omega_n^2$ are restricted from below, shifted to start from $\omega_0^2=0$,  the perturbations constitute a complete set of orthogonal eigenfunctions, and the solutions remain stable. For the sake of simplicity, hereafter we take those solutions as an orthonormal basis, i.e.,
\begin{equation}
\bra{\psi_n}\ket{\psi_m}=
\int_{-\infty}^\infty \psi_n(x)^*\psi_m(x)dx=
\left\{\begin{array}{lr}
\delta_{nm} & n,m\text{ discrete}\\
\delta(n-m)& n,m\text{ continous}\\
\end{array}
\right.
\;,
\end{equation}
where the discrete index corresponds to bounded eigenfunctions while the continuous ones
 are associated to scattering eigenfunction with plane wave asymptotic behaviour.

At this point, the natural question is if can consider the inverse problem, i.e., let us construct
a scalar theory with stable stationary solutions starting from a given one-dimensional Schr\"odinger like system of the form
(\ref{Sch}).
 For this purpose, we use equation (\ref{stateq}) to find the operator relation
\begin{equation}
\label{ChainRule}
\frac{d}{d \varphi}=\frac{1}{\sqrt{2 V(x)}}\frac{d}{dx}\;,\qquad
V(x)=V(\varphi(x))\;.
\end{equation}
With this, the Schr\"odinger potential $U(x)$ in (\ref{Sch}) admits the representation
\begin{equation}
U(x)=-\frac{1}{4 V(x)^2}\left(\frac{dV(x)}{dx}\right)^2+\frac{1}{2V(x)}\frac{d^2V(x)}{dx^2}\;,
\end{equation}
or
\begin{equation}
\label{EqV}
-\left(\frac{dV(x)}{dx}\right)^2+2V(x)\frac{d^2V(x)}{dx^2}-4U(x)V(x)^2=0\;.
\end{equation}
With the change of variable $V(x)=g(x)^2$ \cite{AM1}, the last equation reduces automatically to the zero energy Schr\"odinger equation
\begin{equation}
\label{Sch2}
-\frac{d^2g}{dx^2}+U(x)g(x)=0\;,
\end{equation}
and by comparing this equation with (\ref{Sch}), we realize that the identification $g(x)=\gamma \psi_0(x)$ holds, being $\psi_{0}(x)$ the zero frequency solution in (\ref{Sch}) and $\gamma$ is a numerical proportionality constant. In consequence, the scalar part of the field theoretical energy density is independent of field derivatives of our scalar theory, its corresponding stationary solution and  its classical mass are given by
{\footnotesize
\begin{equation}
\label{GenKink}
V(x)=\left(\frac{\psi_0(x)}{\psi_0(0)}\right)^2\,,\qquad
\varphi^\pm(x)=\pm \frac{\sqrt{2}}{\psi_0(0)} \int_0^{x} \psi_{0}(y)dy\;,\qquad
M=\frac{2\nu^2 }{\lambda(\psi_0(0))^2}\bra{\psi_0}\ket{\psi_0}\;,
\end{equation}}
where we have considered the boundary condition $\varphi(0) = 0$ and fixed $\gamma = \psi_0(0)^{-1}\not=0$ to ensure that $V(0) = 1$. These relationships indicate that to achieve stable solutions in a non-linear scalar theory with a finite mass, we need to focus on a one-dimensional Schr\"odinger system with a bounded ground state. Therefore, $\psi_0(x)$ must satisfy the following conditions
\begin{equation}
\psi_0(x) >0 \quad   \{\forall x |\,  \vert x \vert < \infty\} \quad , \quad \lim_{x\rightarrow \pm \infty}\psi_0(x)=0\;\;,\qquad
\bra{\psi_0}\ket{\psi_0}=1.
\end{equation}
We emphasize that the three equations in (\ref{GenKink}) are the ones to obtain the features of the new field theories from Schr\"odinger equation solution, and is the main result of the paper.

Additionally, from the theory of spectral analysis, one has that
the ground state  $\psi_0$ of a one-dimensional Schr\"odinger operator does not have zeros and is non-degenerated \cite{LL}.
 If we also ask for parity invariance, i.e., $U(-x)=U(x)$ we have
\begin{equation}
\psi_0(-x)=\psi_0(x)\;,\qquad
\varphi^\pm(-x)=-\varphi^\pm(x)\;,
\end{equation}
meaning that the $V(\varphi)$ is symmetric respect to $\varphi(0)=0$, and the stationary solution interpolates between two extreme points $-\varphi_v$ and $\varphi_v$, given by
\begin{equation}\label{def_varphi_v}
\varphi_v=\frac{1}{\sqrt{2}\psi_0(0)}\int_{-\infty}^{+\infty}\psi_0(x) dx=\varphi^{+}(+\infty)\;.
\end{equation}
Typically, these extreme points are referred to as the \emph{ground states} of the scalar theory, and as we will see below, they correspond to the zeros of the potential $V(\phi)$. In this context, the functions  $\varphi^\pm(x) $ exhibit \emph{kink or anti-kink nature}, depending on the sign, representing a transition between two vacuums $\pm\varphi_v$.  This function is also the map $\varphi: x \rightarrow \varphi(x)$, transforming the entire real line into the compact interval $(- \varphi_v, \varphi_v)$. To reconstruct the potential of the scalar theory $V(x)$ as a function of $\varphi$, we can use the inverse mapping; however, this only provides information within the restricted interval,
\begin{equation}
\label{Vvarphi}
V(\varphi)= V(x(\varphi))\,, \qquad
|\varphi|<\varphi_v\,.
\end{equation}
We remark that
\begin{equation}
\varphi(x) \subset \phi(t,x) \in {\rm Maps}(\mathbb{R}^{1,1}, (-\varphi_v,\varphi_v))\,,
\end{equation}
which allows us to formally replace the explicit kink solution $\varphi(x)$ in (\ref{Vvarphi}) with any arbitrary field configuration $\phi(t,x)$ taking values in the interval $(-\varphi_v, \varphi_v)$. Unfortunately, this method cannot reconstruct the potential outside these boundaries. Despite this limitation, we can analytically confirm that $\pm\varphi_v$ correspond to the unique vacua of the theory within this interval. In first place, by using (\ref{def_varphi_v}), we find
\begin{align}
V(\pm \varphi_v)=\lim_{x\rightarrow \pm \infty}\psi_0(x)^2=0\,.
\end{align}
Secondly, the absence of nodes of $\psi_0(x)$ and its symmetry under reflections anticipates that this is the minimum value  $V(\varphi)$ can reach.

To finish this paragraph, let us define a confining kink. For this purpose, we shall discuss on the nature of the perturbations when they move far away from the center, i.e., let us go deep in the nature of the quantity
\begin{equation}
\eta_\infty^\pm =\lim_{x\rightarrow \pm \infty}\eta(\tau, x)\,.
\end{equation}
According with \cite{TrulFle}, the first mode characterized by the ground state of the Schr\"odinger system is called the translational Goldstone mode. It corresponds to the first derivative of the stationary solution itself, and its existence is due to the spontaneous breaking of the translational invariance by the freedom to set the center of mass of the stationary solution at the origin. It is clear that does not contribute to $\eta_\infty$. Other discrete frequencies are associated with the so-called ``internal" oscillation modes in which the stationary solution undergoes harmonic variations around its center. These modes are also bounded and therefore they do not contribute to $\eta_\infty$ as well. Finally, frequencies in a continuous spectrum are associated with the behaviour of free mesons far away the stationary solution when considering the quantum picture. The corresponding modes are not bounded, so they can contribute to $\eta_\infty$.

The asymptotic behavior of the perturbations also influences the potential $V(\phi)$ outside the interval $(-\varphi_v,\varphi_v)$ in the general case. To understand this, let us consider the Taylor expansion
\begin{equation}
	\begin{aligned}
		V(\varphi^{\pm}(x)+\alpha\eta(\tau,x) )&= V(\varphi(x))+\alpha \eta(\tau,x)\frac{\partial V(\phi)}{\partial \phi}|_{\phi=\varphi}+\frac{\alpha^2 \eta(\tau,x)}{2}\frac{\partial V(\phi^\pm)}{\partial^2 \phi^2}|_{\phi=\varphi^\pm}\\
		&=\left(\frac{\psi_0(x)}{\psi_0(0)}\right)^2
		\pm {\sqrt{2}}\,\alpha  \eta(\tau,x) \frac{\psi_0'(x)}{\psi_0(0)}+\frac{\alpha ^2\eta (\tau,x) ^2}{2}U(x)+\ldots\;,
	\end{aligned}
\end{equation}
where in the second line we have explicitly used the equations of motion along with the definitions of $U(x)$ and $\varphi^\pm(x)$ given in (\ref{Sch}) and (\ref{GenKink}), provided that $\alpha \eta(\tau,x)$ is a dimensionless quantity. Then, taking the limit $x\rightarrow \infty$, we obtain
\begin{equation}
	V(\varphi_v+\alpha\eta_\infty)= \lim_{x\rightarrow \infty}\left(	\frac{\alpha^2  \eta(\tau,x)^2}{2} U(x)+\ldots\right)\;.
\end{equation}
This relationship implies that in order for the left-hand side to have a finite value, the product $\eta^2(\tau, x) U(x)$ must also remain finite as $x$ approaches infinity. In typical examples, such as the $\Phi^4$ theory or the sine-Gordon model, these conditions lead to bounded and scattering behavior for the perturbations. With this understanding, we define a ``confining kink" as a scalar theory that meets the following  conditions
\begin{equation}
	\label{BC}
	\lim_{x\rightarrow \pm \infty} U(x)=\infty\;,\qquad  \lim_{x\rightarrow \pm \infty} \eta(\tau,x)=0\;,\qquad
	\lim_{x\rightarrow \infty}	\eta^2(\tau,x) U(x)=0\;.
\end{equation}
In general these conditions imply that
\begin{equation}
	\label{BC0}
	\psi_n(x)|_{x\rightarrow \pm \infty}=0 \qquad \forall n\in \mathbb{N}_0\;,
\end{equation}
where $\mathbb{N}_0$ is the set of all natural numbers, including $0$.

The first equation (\ref{BC}) indicates that the curvature of $V(\phi)$ is infinite in the point $\varphi_v$. As a result, perturbations are not allowed at the exact vacuum point, which represents a cusp. This means that the solutions of the scalar theory are completely restricted to the open interval $(- \varphi_v, \varphi_v)$.

These conditions have not been considered in the study of scalar theories until now, as far as we know, and they will have important consequences at the quantum level.

\subsection{One loop quantum corrections of the kink mass at rest}
\label{SecQuan}

For the quantization of  standard  scalar field models featuring kinks, such as the $\Phi^4$, theory or the sine-Gordon model, it is mandatory a parallel analysis of vacuum fluctuations \cite{TrulFle, AM1, AM2, AMP}. The ultraviolet divergences due to kink fluctuations, or mesons moving in kink backgrounds are tamed, even cancelled, by subtracting the divergences induced by vacuum fluctuations, or mesons travelling in homogeneous backgrounds. However, in the systems we intend to construct, the situation is different. In this section we will see that in ur models there is no need of any subtraction to control ultraviolet divergences, only
$\zeta$-function regularization is sufficient.


To compute the one-loop kink energy Casimir energy, we follow the usual second quantization prescription, i.e., we take the expansion
\begin{equation}
	\hat{\phi}(x,\tau)= \varphi(x)+\sqrt{\hbar}\frac{\lambda}{\nu}\hat{\eta}(\tau,x)\;,
\end{equation}
where $\hat{\eta}(\tau,x)$ describe quantum corrections around the kink.  Plugging this expansion in the field Hamiltonian we obtain
\begin{align}
\label{Quantization}
	\hat{\mathcal{H}} \approx & M+\frac{\hbar \lambda}{2}\int dx \left\{\left(\frac{\partial \hat{\eta}}{\partial\tau}\right)^2+\hat{\eta}\left(-\frac{d^2}{dx^2}+U(x)\right)\hat{\eta}\right\} \;.
\end{align}
where $M$ is the kink classical mass and the other summand is the one-loop kink mass shift proportional to $\hbar$.
At this point, it is worth remarking that systems with confining kinks are defined by the conditions (\ref{BC}) and (\ref{BC0}). Consequently, the spectrum of the operator appearing in the second term of equation (\ref{Quantization}) is purely discrete. Then, by expanding the field operator $\hat{\eta}(\tau,x)$ in terms of bosonic ladder operators $\hat{a}_n$ and $\hat{a}_n^\dagger$ and the perturbation modes $\psi_{n}(x)$ as follows
\begin{equation}
	\hat{\eta}(\tau,x)=\sum_{n}\frac{1}{\sqrt{2\omega_n }}(\hat{a}_ne^{\Ti   \omega_n \tau}+\hat{a}_n^\dagger e^{-\Ti   \omega_n \tau}) \psi_n(x)\;,
\end{equation}
we finally get
\begin{equation}
	\hat{\mathcal{H}}_0=M+\hat{Q}\;,\qquad \hat{Q}=\frac{\hbar \lambda}{2}\sum_{n}\omega_n(2\hat{a}_n^\dagger \hat{a}_n+1)\;.
\end{equation}
We stress that this $\hbar$-expansion around the kink $\varphi(x)$ is perfectly regular at one-loop order provided $U(s)$ is regular on the open real line $x\in (-\infty,\infty)$. The WKB regime completely fails at the boundary points $\phi(\pm \infty)$  if these values are continued in the interior of $x\in (-\infty, \infty)$ because then $U(x)$ becomes $U(\pm \infty)$ which in our models is infinity. Therefore, we envisage field theoretical models where the quantum vacuum sector does not exist even in the WKB approximation. Nevertheless, there exists a bona fide quantum kink sector. The situation is similar to the structure of Liouville field theory  \cite{DHoker1982}, where there is no vacuum sector because the minimum of $V(\phi)\propto e^\phi$ occurs at $\phi=-\infty$. However, a  honest space of quantum states can be constructed over the (singular) kink $\varphi(x) =-\ln(m^2 x^2)$, where $m$ is the parameter of dimension $L^{-1}$ in the corresponding Lagrangian of the model.

 Thus, the one-loop confinig kink mass corresponds to the vacuum expectation value of the operator $\hat{Q}$
in absence of quantum kink fluctuations, i.e. in the state $\hat{a}_n \vert 0 \rangle =0, \forall n$,
\begin{equation}
	\label{Q}\bra{0}\hat{\mathcal{H}}_0\ket{0}=M +Q\;,\qquad Q=\frac{\hbar \lambda}{2}\sum_{n=1}^\infty\omega_n\;.
\end{equation}
As we might expect, the sum over all frequencies in the quantity $Q$ diverges. However, unlike the other cases where scattering states appear, here the sum is always discrete. To regularize it, we can construct the spectral Riemann zeta function \cite{Zeta} as follows:
\begin{equation}
	\zeta_\omega(s) =\sum_{n}\frac{1}{\omega_n^s}\;,\qquad Q(s)=\frac{\hbar \lambda}{2}\zeta_\omega(s)\;.
\end{equation}
As long as the function \(\zeta_\omega(s)\) can be analytically continued to the entire complex plane, allowing for the assignment of a numerical value $Q(s)$ at $s = -1$, the quantum correction to the mass is given by $Q = Q(-1)$. Additionally, at the physical point \(s = -1\), \(\zeta_\omega(s)\) will yield a finite value, resulting in a finite quantum correction without the need for any renormalization schemes.

In the following subsection we apply the entire formalism to build the first example of confining kink models.

\section{The error kink}
\label{ErrorSec}
To construct our first confining kink, we begin by addressing the Schr\"odinger problem (\ref{Sch}) that the corresponding perturbation modes satisfy.

\subsection{Potential, eigenfunctions and kink mass}
\label{ErrorSubSec}
In accordance with the conditions (\ref{BC}), one could naturally consider a shifted harmonic oscillator, i.e.,
\begin{equation}
	\label{harmonic}
	\left( -\frac{d^2}{d x^2}+ U(x)\right) \psi_n(x)=2 n \psi_n(x)\;,\qquad
	U=x^2-1\;,\qquad n=0,1,\ldots,
\end{equation}
where the solutions are given in terms of the Hermite polynomials $H_n(x)$ as follows \cite{Flugge},
\begin{equation}
	\psi_n(x)=\frac{H_n(x)e^{-\frac{x^2}{2}}}{\pi^\frac{1}{4}\sqrt{ 2^n n!}}\;,\qquad
	\bra{\psi_n}\ket{\psi_m}=\delta_{nm}\;.
\end{equation}
By replacing in formulas given in  (\ref{GenKink}) for $V(x)$, the stationary solution, and its respective mass, we get
\begin{eqnarray}
	\label{intermsx}
	V(x)= e^{-x^2}\;,\qquad
	\varphi(x)=\sqrt{\pi } \, \text{erf}\left(\frac{x}{\sqrt{2}}\right)\;,\qquad
	M_0=\frac{2 \sqrt{\pi}\nu^2 }{\lambda}\;,
\end{eqnarray}
where error function
\begin{equation}\label{err_func}
	\text{erf}(x) = \frac{2}{\sqrt{\pi}} \int_0^x e^{-t^2} dt \;,
\end{equation}
has a domain that covers the entire real line, while its range is limited to the interval \((-1, 1)\). The stationary solution \( \varphi(x) \) takes the form of a kink that interpolates between two vacuum states, denoted as \( \varphi_v = \pm \sqrt{\pi} \); we refer to this as the \emph{error kink}. Conversely, all perturbation modes are bounded, and they disappear at $x\rightarrow \infty$.

By using the error kink, one has that the potential of the corresponding scalar theory in terms of the field
$\varphi\rightarrow \phi$ is given by
\begin{equation}
	\label{V(f)error}
	V(\phi)=\left\{\begin{array}{lc}
		\exp(-2\, \left[\text{erf}^{-1}\left(\frac{\phi}{\sqrt{\pi}}\right)\right]^2)&|\phi| <\sqrt{\pi}\\
		0&|\phi| =\sqrt{\pi}\\
		\infty&|\phi| >\sqrt{\pi}\\
	\end{array}\right.\;,
\end{equation}
where $\text{erf}^{-1}(x)$ is the inverse of the error function with domain and range are $(-1,1)$ and $\R$, respectively. The complementation of the potential with infinite walls located at the vacuum points $|\varphi_v|=\sqrt{\pi}$ is quite natural to assume (as we do for all models in this article) since the curvature of the potential goes to infinity at these points,
\begin{align}
	&\frac{\partial V(\phi)}{\partial \phi}=-2 e^{-\text{erf}^{-1}\left(\frac{\phi}{\sqrt{\pi }}\right)^2} \text{erf}^{-1}\left(\frac{\phi}{\sqrt{\pi }}\right) &&\Rightarrow &&
	\frac{\partial V(\phi)}{\partial \phi}|_{\phi=\pm\sqrt{\pi}}=0\;.
	\\\label{Secondder}
	&\frac{\partial^2 V(\phi)}{\partial \phi^2}=
	2 \text{erf}^{-1}\left(\frac{\phi}{\sqrt{\pi }}\right)^2-1 &&\Rightarrow &&
	\frac{\partial^2 V(\phi)}{\partial \phi^2}|_{\phi=\pm\sqrt{\pi}}=+\infty.
\end{align}

Moreover, it is possible to show that the second derivative near the vacuum diverges logarithmically as $\phi\rightarrow \pm\sqrt{\pi}$, i.e.,
\begin{equation}
\lim\limits_{\phi\to\sqrt{\pi}^{-}} \frac{\partial^2 V(\phi)}{\partial \phi^2} = \lim\limits_{\phi\to-\sqrt{\pi}^{+}} \frac{\partial^2 V(\phi)}{\partial \phi^2} = \ln \left(\frac{2}{\left(\sqrt{\pi }-\phi \right)^2}\right)+\ln \left(\ln \left(\frac{2}{\left(\sqrt{\pi }-\phi \right)^2}\right)\right)-1
\;.
\end{equation}
This result is the geometric manifestation that the mass of the fundamental quanta $U_\infty=U(\sqrt{\pi})$ in systems of this type is infinite. The curvature of the potential evaluated in the homogeneous ground state is precisely the mass of the meson particle. However, evaluated at the error kink $\varphi=\sqrt{\pi}{\rm erf}\left( \frac{x}{\sqrt{2}}\right)$, the second-order derivative of the field-theoretic potential gives rise to the quantum-mechanical harmonic oscillator potential,
\begin{equation}
	\frac{\partial V(\phi)}{\partial \phi}|_{\phi=\pm \varphi}= -\sqrt{2} x e^{-\frac{x^2}{2}}
	\;,\qquad
	\frac{\partial^2 V(\phi)}{\partial \phi^2}|_{\phi=\varphi}=x^2-1=U(x)\;.
\end{equation}

\begin{figure}[H]
	\begin{center}
		\includegraphics[scale=0.55]{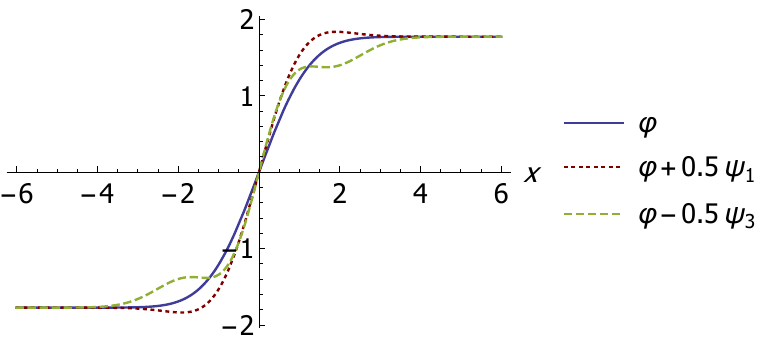}
		\includegraphics[scale=0.48]{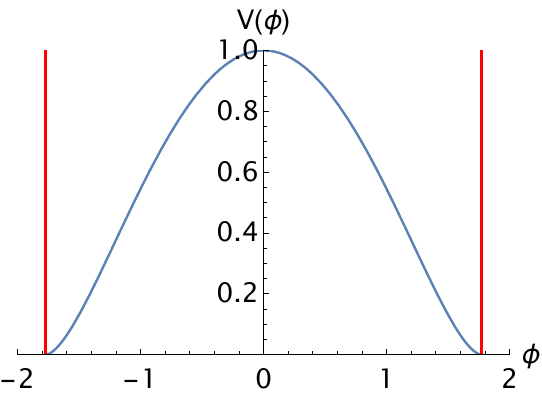}
	\end{center}
	\caption{\small Left panel: Plot of the error kink and some perturbation modes. Here we use the perturbation parameter $\alpha=0.5$ to exaggerate the effect of the deformation. Right panel: Plots of the potential $V(\phi)$.}
	\label{figErr}
\end{figure}

In Figure \ref{figErr} we plot the error kink along with some perturbation modes, as well as the shape of the model potential.

At this point is it worth to mention that, for $\phi\sim 0$, this by adding an unessential shift of $\frac{1}{2}$, potential takes the form of the potential for the $\phi^4$ theory
\begin{equation}
	V(\phi)\approx \frac{3}{2} \left(1-\frac{\phi ^2}{6}\right)^2+\mathcal{O}(\phi^6)\;.
\end{equation}
Accordingly, this result shows that the error function can be accurately approximated by a series of powers of hyperbolic tangents; see \cite{Basset,Reyes}. Therefore, one may consider the error kink as a nonlinear superposition of kinks. This argument will be a key to interpreting the result at the quantum level in terms of the WKB expansion in the following subsection.

Let us compute the mass of the error kink by using only the explicit form of the potential, i.e., with the help of formula  (\ref{MW}). This gives us
\begin{equation}
	M_0=\frac{\nu^2}{\lambda} \int_{-\sqrt{\pi}}^{\sqrt{\pi}} d\varphi \sqrt{2 V(\varphi)} =\frac{\nu^2}{\lambda} \sqrt{\pi } \text{erf}\left(\sqrt{2} \text{erf}^{-1}\left(\frac{\varphi}{\sqrt{\pi }}\right)\right)\Big\vert_{-\sqrt{\pi}}^{\sqrt{\pi}}=\frac{2 \sqrt{\pi}\nu^2 }{\lambda}\;,
\end{equation}
which is the same result as the third equation in  (\ref{intermsx}).

We refer to Section \ref{SecQuan} to canonically quantize the system. Following the arguments that lead to the generic DHN formulation, see \cite{AM2}, we find that in the case of the error kink, only the contribution from the bound states remain.  Moreover,  the spectrum of the Schrödinger operators with the potentials  $U(x) = x^2 - 1 $ and $ U_\infty$ does not include scattering states at all. Consequently, the regularized one-loop DHN mass shift derived from the spectral $\zeta$ function
\begin{equation}
	\label{Errorz}
	\zeta_\omega(s)=\sum_{n=1}^\infty \frac{1}{(2n)^\frac{s}{2}}=\frac{1}{2^{\frac{s}{2}}}\zeta(s/2)\;,
\end{equation}
where $\zeta(s)$ is the Riemann zeta function,  has a finite numerical value at the physical point $s=-1/2$. Therefore, the one-loop quantum correction of the error kink mass is given by
\begin{equation}
	\label{Qerror}
	Q(-1)=\frac{\hbar \lambda}{\sqrt{2}}\zeta(-1/2)\approx-0.147\lambda \hbar\;.
\end{equation}
It is worthwhile to recall that the zero mode $\psi_0(x)=e^{-\frac{x^2}{2} }$ does not contribute to the kink mass shift at one-loop order.
Similar to the case of the standard Casimir effect, quantum corrections are negative \cite{Bordag2001}. In analogy with such effect,  (\ref{Qerror}) may imply that there is an attractive force pointing  out to the center. Although the trapping of all perturbations is mainly due to the confining potential $U(x)= x^2-1$ the semi-classical attractive force contributes mildly to this effect.

\subsection{WKB analysis in the error model}

In the previous subsection the main features of the scalar field theoretical model that we shall term as \lq\lq error\rq\rq have been described through the inverse approach. A deeper analysis about how the WKB approximation work or not work
in this model requires a direct attack. We start now from the classical Lagrangian density

\begin{eqnarray*}
\cal{L}(\phi,\partial_\mu\phi)&=& \frac{1}{2}\left(\frac{\partial\phi}{\partial\tau}\right)^2- \frac{1}{2}\left(\frac{\partial\phi}{\partial x}\right)^2- V[\phi(x)]\,,
\\ V[\phi(x)]&=& {\rm exp}\left[-2\left({\rm erf}^{-1}[\frac{\phi(x)}{\sqrt{\pi}}]\right)^2\right]\,.
\end{eqnarray*}

The $\hbar$-expansion of the part of the energy density without field derivatives in the kink sector is now written up to the fourth-order in $\sqrt{\hbar}$
\begin{eqnarray}
V[\sqrt{\pi} \erf[\frac{x}{\sqrt{2}}]&+&\sqrt{\hbar}\eta(\tau,x)] \simeq  e^{-x^2}-\sqrt{2 \hbar} \, x\,  e^{-\frac{x^2}{2}}\, \eta(\tau,x)+ \frac{\hbar}{2}(x^2-1) \eta^2(\tau,x)+ \\ &+& \frac{(\sqrt{\hbar})^3}{3 \sqrt{2}}\, x \, e^{\frac{x^2}{2}} \eta^3(\tau,x)+\frac{\hbar^2}{24}\, (1+x^2)\, e^{x^2} \eta^4(\tau,x) + {\cal O}(\hbar^{\frac{5}{2}}) \nonumber
\end{eqnarray}
The asymptotic expansion has been done by performing all the functional derivatives at coinciding space-time points chosen at a distance $x$ of the kink center. The dangers that this choice could produce in the full field theory may be tamed for instance through a point-splitting regularization. It may be checked that the term proportional to $\sqrt{\hbar}$ in the formula above is exactly canceled by the term $-\sqrt{\hbar}\frac{d^2\varphi}{d x^2}$ arising from the kinetic terms. The $\hbar$ term gives the Schr\"odinger
potential of the harmonic oscillator that we took as starting point in the previous subsection. Of course it must be added to the the terms of the same order coming from the kinetic terms. Finally, point-dependent trivalent and fourvalent vertices exponentially growing with the distance to the kink center arise at the next orders. The couplings of growing strength are attractive and this fact reinforces the confining character of the error kink.

Trying the WKB expansion in the vacuum sector we find the exact result
\begin{equation}\label{expansion_vacuum}
V[\sqrt{\pi} +\sqrt{\hbar}\eta(\tau,x)] =   \frac{1}{2}\hbar \eta(\tau,x)^{2} \ln(\frac{2}{\hbar\,\eta(\tau,x)^{2}})\;,
\end{equation}
meaning there are no iterated higher-order logarithm terms. This lack of Taylor expansion was somehow expected; trying any power series around the vacuum, convergent or asymptotic, is nonsense because the second derivative of the potential at that value of the field is infinite, as can be read from (\ref{Secondder}). The logarithmic structure in equation (\ref{expansion_vacuum}) indicates that the WKB expansion around the vacuum sector is not a conventional asymptotic power series in $\hbar$, but instead is based on a non-analytic branch point often appearing in trans-series, implying that the system exhibits a complete non-perturbative behaviour in the vacuum sector. Consequently, the error model confines all physically consistent quantum information exclusively to the kink sector. This characteristic, in which the error model and related models discussed in the following section lack well-defined vacuum sectors but possess genuine kink sectors, is not unprecedented. d’Hoker and Jackiw previously demonstrated that the Liouville and Toda field models exhibit similar properties \cite{DHoker1982}, although in those cases the ground state is located at $\phi_v = -\infty$. In the present family of models, this arises because the potential energy density displays infinite discontinuities in its first derivative at the ground states.

The part of the energy density with no field derivatives in the error model may be understood  as the limit of the truncated power series expansion of $V[\phi]$ around $\phi=0$,
\begin{equation}\label{V_phi_N}
V[\phi,N]= \sum_{j=0}^N \, \frac{1}{j!} \, \frac{d^j V}{d \phi^j}\big\vert_{\phi=0} \, \phi^j \quad , \quad V[\phi]=\lim_{N\to \infty} V[\phi,N] \;.
\end{equation}

For finite $N$ all of these models are completely well defined. For instance, the features of the lower $N=4$ case are as follows
\begin{eqnarray*}
&& V[\phi,4]=1-\frac{\phi^2}{2}+\frac{\phi^4}{24} \quad , \quad V[0,4]=1 \quad , \quad \varphi^{(4)}_v=\pm\sqrt{6}\;, \\
&&  V[\pm \sqrt{6},4]=-\frac{1}{2}\;,  \frac{d^2 V}{d \phi^2 }\Big\vert_{\varphi_v}=\lim_{x \to \pm\infty} U(x,4)=2 \;.
\end{eqnarray*}
Clearly, in this case, the $N=4$  system admits a WKB expansion in both the vacuum and the kink sector, so the DHN formula also incorporates zero-point and mass renormalization. Increasing $N$ in an even integer, the new function still exhibits two minima displaced towards the origin until reaching $\varphi_v=\pm \sqrt{\pi}$ at $N=\infty$. Also, the absolute value at the minima increases up to zero in the $N\to\infty$ limit, and the threshold of the continuous spectrum in $U(x,N)$ remains finite but rapidly increasing with $N$ as long as $N$ does not reach $\infty$.

\begin{figure}[H]	
\begin{center}
\includegraphics[scale=1.1]{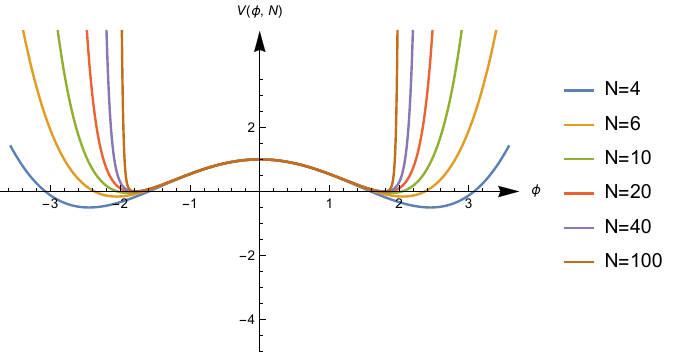}
\end{center}
\caption{\small The potential $V[\phi,N]$ defined in (\ref{V_phi_N}) for different values of $N$. It can be seen when $N\to+\infty$, then the plot of $V[\phi,N]$ tends to the right-panel plot of Figure \ref{figErr}.}
\label{fig_VN}
\end{figure}

To test these statements let us briefly consider the $N=10$ case
\begin{eqnarray*}
V[\phi,10]&=& 1-\frac{\phi^2}{2!}+\frac{\phi^4}{4!}+2\frac{\phi^6}{6!}+13 \frac{\phi^8}{8!}+88 \frac{\phi^{10}}{10!}\;, \qquad V[0,10]=1 \\ \varphi_v&\simeq& \pm 1.89\;, \quad V[\varphi_v,10]\simeq-0.05 \;, \frac{d^2 V}{d \phi^2 }\Big\vert_{\varphi_v}=\lim_{x \to \pm\infty} U(x,10)\simeq3.36\;,
\end{eqnarray*}
and the $N=100$ case
\begin{eqnarray*}
 && \qquad V[0,100]=1\;, \varphi_{v}\simeq\pm 1.78 \;, V[\varphi_v,10]\simeq-0.000366 \;, \\
  &&\mbox{and} \quad \frac{d^2 V}{d \phi^2 }\Big\vert_{\varphi_v}=\lim_{x \to \pm\infty} U(x,100)\simeq7.12\;.
\end{eqnarray*}
These data are obtained using numerical methods since analytical kink solutions are not available in this case. In Figure \ref{fig_VN}, it is shown how when $N\to+\infty$, $V[\phi,N]\to V(\phi)$, and $\varphi_{v}\to\pm\sqrt{\pi}$. Also, notice that $V[\varphi_{v}]\to0^{-}$ when $N\to+\infty$.

The error kink is a specific example of a confining kink. It raises the question of how to develop theories that exhibit similar characteristics, such as all bounded perturbations and $\zeta$-regularized finite quantum corrections. In the next section, we will explore a formalism that enables us to create an infinite family of models with comparable properties while also incorporating essential differences.

\section{New species of confining kinks}
\label{sec_new_species}

In this section, we rely on the family of rationally deformed harmonic oscillators to build models that exhibit localized fluctuations around the kink solitary waves. Such rational deformed harmonic oscillators are constructed using the Darboux transformation, which is detailed in the Subsection \ref{sec_DarbouxT}. As a result, we obtain new scalar theories whose stationary solution only admits bounded perturbations over the kink in Subsection \ref{sec_Owen}. In other words, the system traps all quantum kink perturbations. Moreover, the lowest frequency of vacuum state perturbations is infinite, and fundamental quanta do not propagate. Consequently, the quantum correction of the systems will not need any ultraviolet renormalization. Although, $\zeta$-function regularization will render one-loop kink mass shifts finite.

\subsection{Darboux transformation}
\label{sec_DarbouxT}

Darboux transformation is a technique that comes from the theory of supersymmetric quantum mechanics and non-linear integrable differential equations  \cite{Cooper,Matveev}. It is mostly used to construct new Schr\"odinger-like problems based on well-known models. It has been also considered in the study of quantum correction on scalar field theories associated with P\"oschl-Teller quantum systems \cite{AMP}. In this article, we will utilize this method to derive Schr\"odinger Hamiltonian operators with a discrete spectra that are bounded from below. The corresponding solutions will be interpreted as the perturbation modes of the stationary solution of a scalar theory, as described in equation (\ref{Sch}). To explain the method, we will start with a stationary Schr\"odinger equation of the following form:
\begin{equation}
	\label{originsys}
	H_0\psi_{\epsilon}=\epsilon\psi_\epsilon\;,\qquad
	H_0=-\frac{d^2}{dx^2}+U(x)\;,
\end{equation}
where the solutions $\psi_\epsilon$ for any eigenvalue $\epsilon$ are known. From these solutions, we select a \emph{``seed state,"} denoted as \(\chi_1\) (where \(H_0\chi_1 = \varepsilon_1 \chi_1\)), to distinguish it from the others. This state serves to build the intertwining operators
\begin{equation}\label{inter_operators}
	A_1=\frac{d}{dx}-\frac{d}{dx}\ln(\chi_1)\;,\qquad
	A_1^\dagger=-\frac{d}{dx}-\frac{d}{dx}\ln(\chi_1)\;,
\end{equation}
which satisfy
\begin{equation}
	\label{Factor}
	A_1^\dagger A_1=H_0-\varepsilon_1\;,
	\qquad
	A_1 A_1^\dagger =H_1-\varepsilon_1\;.
\end{equation}
In the last equation, the new operator $H_1$ is a deformation of the original one, and its explicit form is
\begin{equation}
	\label{firstOrdDT}
	H_1=-\frac{d^2}{dx^2}+U(x)-2\frac{d^2}{dx^2}\ln(\chi_1)\;.
\end{equation}
From here we note that if we do not want a divergent system, the state $\chi_1$ should be nodeless, which strongly restrict our election to the ground state $\chi_1(x)=\psi_{0}(x)$.

Equations (\ref{Factor}) imply the so-called intertwining relations
\begin{equation}
	\label{inter1}
	A_{1}\,H_0=H_{1}\,A_{1}\;,\qquad
	A^\dagger_{1}\,H_{1}=H_{0}\,A^\dagger_{1}\;,
\end{equation}
which establishes a map between the states of both systems. As result we got that
\begin{equation}
	H_{1}\psi_\epsilon^{(1)}=\epsilon \psi_\epsilon^{(1)} \;,
	\qquad
	\psi_\epsilon^{(1)}=\frac{1}{\sqrt{|\epsilon-\varepsilon_{1}|}}A_1\psi_\epsilon\;.
\end{equation}
where the constant factor in the definition of $\psi_\epsilon$ ensures that
\begin{equation}
	\bra{\psi_\epsilon^{(1)}}\ket{\psi_\epsilon^{(1)}}=\frac{1}{|\epsilon-\varepsilon_1|}\bra{ \psi_\epsilon}A^\dagger A\ket{\psi_\epsilon}=\bra{\psi_\epsilon}\ket{\psi_\epsilon}\;.
\end{equation}

It is important to note that  $A_1$ annihilates $\chi_1$, meaning this state cannot be transformed into an eigenvector of the new system. Instead, we recognize that the function $1/\chi_1$ is annihilated by $A_1^\dagger $, making it the ``missing state" associated with the eigenvalue $\varepsilon_1$ for $H_1$ \cite{Cooper}. Now, let us assume that $\chi_1$ is a bound state that vanishes at infinity. If that is the case, the missing state will diverge, making it an inadmissible physical solution. As a consequence, the energy level $ \varepsilon_1$ will be absent in the spectrum of the new system.

The procedure described above can be iterated as much as we want. Before fixing any specific regularity conditions, for $n$ steps we must chose a set of $n$ different but arbitrary seed states
\begin{equation}
	\Delta=\{\chi_1,\ldots,\chi_n\}\;,
\end{equation}
and the resultant system will have the form  \cite{Cooper,Matveev}

\begin{equation}
	\label{Hdelta}
	H_{\Delta}=-\frac{d^2}{dx^2}+U_{\Delta}\;,\qquad U_{\Delta}=U(x)-2\frac{d^2}{dx^2}\ln(W(\Delta))\;,
\end{equation}
where $W(f_1,\ldots,f_n)$ is the generalized Wronskian of $n$ functions,
\begin{equation}
	W(f_1,\ldots,f_n)=
	\begin{array}{|cccc|}
		f_1 & f_2 &\ldots, &f_n\\
		f_1'& f_2'& \ldots&f_n'\\
		\vdots&\vdots&\ddots &\vdots\\
		f_{1}^{(n-1)} &f_{2}^{(n-1)}&\ldots & f_{n}^{(n-1)}\;.
	\end{array}
\end{equation}
To produce non-singular potentials with  this higher order transformation, instead of asking for nodeless seed states, we must take care that this Wronskian does not vanish in any allowed point. This requirement is less strict to its analogue of the first-order case where only the ground state was allowed, and actually, it depends winch kind of system we want to construct. In this article we use the Adler-Krein theorem criteria \cite{Krein, Adler} which will be detailed in the following section. 

As in the first-order case, in the general case we also have an intertwining relation which is given by
\begin{equation}\label{interAAA}
	\mathbb{A}_nH_0=H_{\Delta}\mathbb{A}_n\;,\qquad
	\mathbb{A}_n^\dagger H_\Delta=H_{0}\mathbb{A}_n^\dagger\;,
\end{equation}
where $\mathbb{A}_n$ and  $\mathbb{A}_n^\dagger$ are the accumulated intertwining operator $(A_0=1)$
\begin{align}\label{inter2}
	\mathbb{A}_n&=A_n A_{n-1}\ldots A_1\;,\qquad \mathbb{A}_n^\dagger=A_1^\dagger \ldots A^{\dagger}_{n-1}A_n^\dagger\;, \\
	A_{i}&=\frac{d}{dx}-\frac{d}{dx}\ln(\mathbb{A}_{i-1}\chi_i)\;, \quad A^{\dagger}_{i}=-\frac{d}{dx}-\frac{d}{dx}\ln(\mathbb{A}_{i-1}\chi_i)\;,
\end{align}
which satisfy
\begin{equation}\label{inter3}
	\mathbb{A}_n^\dagger \mathbb{A}_n=\prod_{i=1}^{n}(H_0-\varepsilon_i)\;,\qquad
	\mathbb{A}_n\mathbb{A}_n^\dagger =\prod_{i=1}^{n}(H_\Delta-\varepsilon_i)\;.
\end{equation}
In correspondence with formula (\ref{inter2}), operator $\mathbb{A}_n$ transforms the state of the original system in the state of the new one as follows,
\begin{equation}
	\label{map}
	\psi_\epsilon^{(\Delta)}=[\Pi_{i=1}^{n}|\epsilon-\varepsilon_i|]^{-\frac{1}{2}}\mathbb{A}_n \psi_\epsilon=[\Pi_{i=1}^{n}|\epsilon-\varepsilon_i|]^{-\frac{1}{2}}\frac{W(\Delta,\psi_\epsilon)}{W(\Delta)}\;,
	\qquad
	H_\Delta \psi_\epsilon^{(\Delta)}=\epsilon \psi_\epsilon^{(\Delta)}\;,
\end{equation}
for any $\epsilon\not=\varepsilon_i$ while the set of seed states are annihilated.
By extending the reasoning for the missing states of the first-step case, it is easy to conclude that if the seed states are bounded, their corresponding energy levels will be absent in the spectrum of $H_\Delta$, i.e.,
\begin{equation}
	\text{spectrum}(H_\Delta)=\{\epsilon^\Delta\}=\{\epsilon\}-\{\varepsilon\}\;.
\end{equation}

The possibility of removing energy levels at will allows us to design a variate set of system with non-trivial spectrum. Such systems will be associated with the concept of algebraic intractability \cite{Cha}, and hidden (super)symmetries can appear
\cite{Carinena:2017zfy,Inzunza:2018jdt,Inzunza:2019xml}.

In the following sections we use this formalism to construct new confining scalar potentials within its corresponding one-loop corrections.

\subsection{The Owen kink}
\label{sec_Owen}
Following the same logic as the subsection \ref{ErrorSubSec}, we now construct new exotic theories with confining kinks. For this purpose, we utilize the Darboux transformation formalism described in section \ref{sec_DarbouxT} to create a rational deformation of the harmonic oscillator. 
 The prescription to procedure non-singular deformation corresponds to the Krein-Adler theorem \cite{Krein, Adler}, which tells us that the Wronskian of pairs of subsequent bound states will not vanish in the real line.

To warm up, let us consider the simplest case
\begin{equation}
\Delta_1= \{\psi_1, \psi_2\}=(1,2)\;,
 \end{equation}
 which according with equation (\ref{Hdelta}), produces the following rationally deformed harmonic oscillator Hamiltonian
 \begin{equation}
 \label{H(12)}
 H_{(1,2)}=-\frac{d^2}{dx^2}+U_{(1,2)}\;,\qquad U_{(1,2)}=x^2+\frac{8 \left(2 x^2-1\right)}{\left(2 x^2+1\right)^2}+3\;.
\end{equation}

Since the transformation is of order two, we have a second order accumulated intertwining operator
\begin{equation}
\mathbb{A}_{(1,2)}=\left(\frac{d}{dx}+\frac{4 x}{2 x^2+1}-x-\frac{1}{x}\right)\left(\frac{d}{dx}+\frac{1}{x}-x\right)\;,\qquad
\ker{\mathbb{A}_{(1,2)}}=\{\psi_1, \psi_2\}\;,
\end{equation}
that can be used to map the harmonic oscillator solutions to the states of (\ref{H(12)}), in the sense of intertwining relation (\ref{interAAA}). The ground state
is given by
\begin{equation}
\label{Or0}
\psi_{0}^{(1,2)}=\frac{\sqrt{2}}{\pi^{\frac{1}{4}}}\frac{ e^{-\frac{x^2}{2}}}{2 x^2+1}\;,\qquad
 \omega_0=0\;,\\
\end{equation}
while the exited states are of the form
\begin{equation}
\label{Or2}
 \psi_{n}^{(1,2)}=\frac{\mathbb{A}_{(1,2)}\psi_{n+2}}{2\sqrt{n(n+1)}}=\frac{\sqrt{n (n+1)}}{2 \sqrt{2^{n+1} (n+2)!}} \widehat{H}_{n+2}(x)\psi_{0}^{(1,2)}\;,
\qquad
 \omega_n^{2}=2(n+2)\;,\
\end{equation}
where $n=1,2,\dots\;,$ and
\begin{equation}
\widehat{H}_{l}(x)=4 (l-3) l H_{l-4}(x)+4 l H_{l-2}(x)+H_l(x)\;,\qquad n\not=1,2\;,
\end{equation}
are the first example of exceptional Hermite polynomials of degree $\text{deg}(\widehat{H}_l)=l$ whose weight function corresponds to (\ref{Or0}), see \cite{quesne,quesne2,Gomez}. They are given this name because the degree sequence does not include polynomials of degree $l = 1$ and $l = 2$. This characteristic is also evident in the spectrum, where the distance between the ground  level and the first excited level is twice the distance between any two subsequent levels
\begin{equation}
\omega_1^2 - \omega_0^2 = 2\delta\omega^2\;,\qquad\delta\omega^2 = \omega_{n+1}^2 - \omega_n^2 = 2\;.
\end{equation}
In other words, there appears to be a gap between the isolated ground state and the rest of the spectrum. This system and various deformations have been explored in previous studies \cite{Carinena:2016hfq, Carinena:2017bqs, Carinena:2017zfy}. These works connected the isolated states to the other energy levels using the ABC higher-order ladder operators. Consequently, it follows that any set of isolated energy levels corresponds to reducible representations of non-linear $W$ algebras \cite{Wal1,Wal2}.

\begin{figure}[H]	
\begin{center}
\includegraphics[scale=0.5]{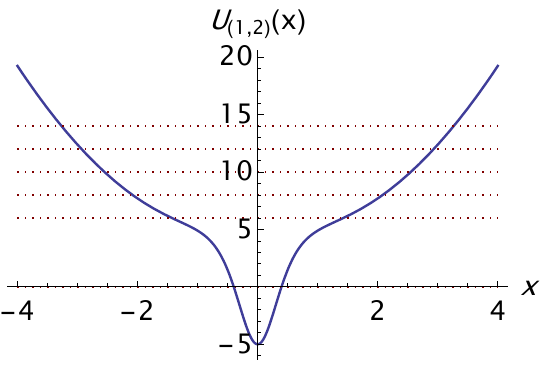}
\end{center}
\caption{\small The Schr\"odinger potential $U_{1,2}$ with its spectrum (dashed straight lines).}
\label{figU0}
\end{figure}

The potential of this operator together with the corresponding allowed energy levels are shown in Figure \ref{figU0}.

\begin{figure}[H]	
\begin{center}
\includegraphics[scale=0.55]{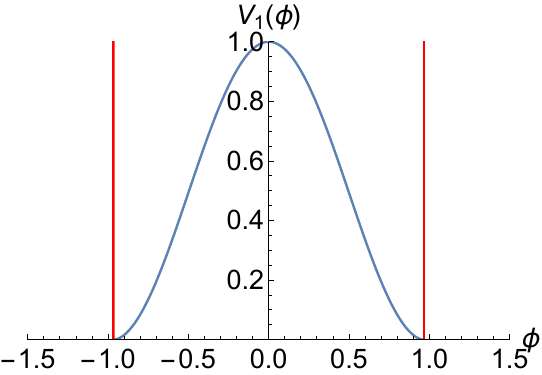}
\includegraphics[scale=0.5]{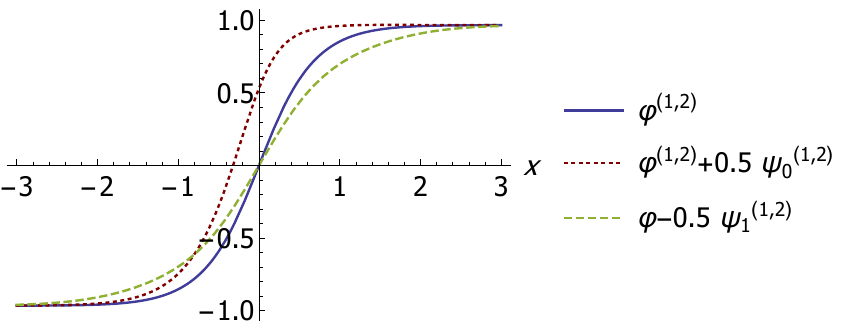}
\end{center}
\caption{\small Left panel: the potential of the scalar theory. Right pane: the Owen kink solution and perturbations.}
\label{fig44}
\end{figure}

By replacing the ground state  $\psi_{0}^{(1,2)}$ on formulas (\ref{GenKink}) we get
\begin{equation}
V(x)= \frac{ e^{-\frac{x^2}{2}}}{2 x^2+1}\;,\qquad
\varphi^{(1,2)}(x)=2 \pi e^{\frac{1}{4}} T\left(\frac{1}{\sqrt{2}},\sqrt{2} x\right)\;,\qquad
 M=\frac{\sqrt{\pi}\nu^2 }{\lambda  }=\frac{1}{2}M_0\;,
\end{equation}
where $M_0$ is the mass of the error kink (\ref{intermsx}), and
the \emph{Owen kink} $\varphi^{(1,2)}(x)$ is given in terms of the
Owen$-T$ function \cite{Owen},
\begin{equation}
T(a,x)=\frac{1}{2\pi}\int_0^x\frac{e^{-\frac{a^2}{2}(1+t^2)}}{1+t^2}dt\;.
\end{equation}
To compute the vacuum of the theory, i.e., the value of the stationary solution at infinity, we note that
\begin{align}
& \frac{\partial}{\partial a}T(a,\infty)=-\frac{1}{2\pi}\int_{0}^{\infty} \, a e^{-\frac{1}{2} a^2 \left(t^2+1\right)}dt=- \frac{e^{-\frac{a^2}{2}}}{2 \sqrt{2 \pi }}\;,
\end{align}
so we can integrate this expression respect to variable $a$ as follows
\begin{align}
T(a,\infty)=\int_a^\infty db \int_0^\infty e^{-\frac{b^2}{2}(1+t^2)}dt=\frac{1}{4} \text{erfc}\left(\frac{a}{\sqrt{2}}\right)>0\;.
\end{align}
where $\text{erfc}(x)=1-\text{erf}(x)$ is the complementary error function. In consequence, the vacuum of the Owen kink theory is given by
\begin{equation}
\varphi_{v}^{(1,2)}=\varphi_{v}^{(1,2)}(\pm \infty)=\pm \frac{\pi}{2} e^{\frac{1}{4}}\text{erfc}\left(1/2\right)\approx \pm  0.967\;.
\end{equation}
Finally, the scalar potential of the new theory is given in terms of the inverse function of $y(x)=T(a,x)$,  denoted as $x(y)=T^{-1}\left(a,y\right)$. Its final form is,
\begin{equation}
\label{OwenP}
V^{(1,2)}(\phi)=\left\{\begin{array}{ll}
\frac{2 \pi ^2 }{T^{-1}\left(\frac{1}{\sqrt{2}},\phi\right)^2+2 \pi ^2 \sqrt{e}}\exp(\frac{1}{2}-\frac{T^{-1}\left(\frac{1}{\sqrt{2}},\phi\right)^2}{8 \sqrt{e} \pi ^2}) \qquad & |\phi|<\varphi_{v}^{(1,2)}\\
0 & \phi=0\\
\infty & |\phi|>\varphi_{v}^{(1,2)}
\end{array}\right.\;,
\end{equation}
In Figure \ref{fig44}, we show the form of the potential of the scalar theory and the stationary solution with its respective perturbations with $\alpha=0.5$.
The quantization of a scalar theory with potential (\ref{OwenP}) is done by constructing the  related  spectral-$\zeta$ function, which yields
\begin{equation}
\zeta_{\omega_n}(s)=\frac{1}{2^{\frac{s}{2}}}\sum_{n=1}^{\infty}\frac{1}{(n+2)^\frac{s}{2}}=\frac{1}{2^{\frac{s}{2}}}\left(
-\frac{1}{1^\frac{s}{2}}-\frac{1}{2^\frac{s}{2}}+\zeta\left(\frac{s}{2}\right)\right)\;,
\end{equation}
where according to Dashen-Hasslacher-Neveu prescription \cite{dhn1,dhn2} the zero mode has been given up to this order. The spectral function of the Owen kink model, can be described as the one of the error kink model (\ref{Errorz}), with a specific factor subtracted due to the Darboux transformation procedure. As a result, this leads to a larger negative quantum correction of the mass when compared to the error kink,
\begin{equation}
Q(-1)=\frac{\hbar \lambda}{\sqrt{2}}\left(
\zeta(-1/2)-1-\sqrt{2}
\right)\approx-1.854\hbar \lambda\;.
\end{equation}

\subsection{Deformed confining kinks}
In the previous paragraph, we considered the first non-trivial rational deformation of the harmonic oscillator. This system features a spectrum with a band gap between the ground and first excited states. In the same vein, the Darboux transformation technique generally allows us to construct a system with multiple gaps of different sizes, positioned wherever we desire. The more complicated the spectrum is, the more intricate the form of the potential and the associated states will be. It is reasonable to expect that these features will manifest somehow in the associated scalar theories and their stationary solutions. In this subsection, we explore this issue. First, we consider the following Darboux schemes for construing new systems:
\begin{equation}
\Delta_{J}=\{\psi_{J},\psi_{J+1}\}=(J,J+1)\;,\qquad
J=1,2,\ldots\;.
\end{equation}
This particular election of the seed states produces the following Schr\"odinger potential
\begin{align}
\label{UJ}
&U_{\Delta_J}:=U_{J}(x)=-\frac{2 P_J''(x)}{P_J(x)}+\frac{2 P_J'(x)^2}{P_J(x)^2}+x^2+3\;,
\\
& P_J(x)=(J+1) H_{J}(x){}^2-J H_{J-1}(x) H_{J+1}(x)\;.
\end{align}
The potential $U_{J}(x)$ does not have singularities on the real line since Hermite polynomials, $H_{j}(x)$, with different indices do not simultaneously vanish at the same point. Note that for the case $J=1$, (\ref{UJ}) reduces to the potential term in  (\ref{H(12)}), so for this particular value, we recover the Owen kink theory.

\begin{figure}[H]
\begin{center}
\includegraphics[scale=0.5]{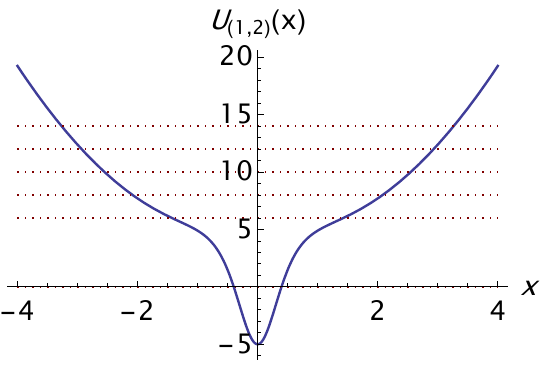}\quad
\includegraphics[scale=0.5]{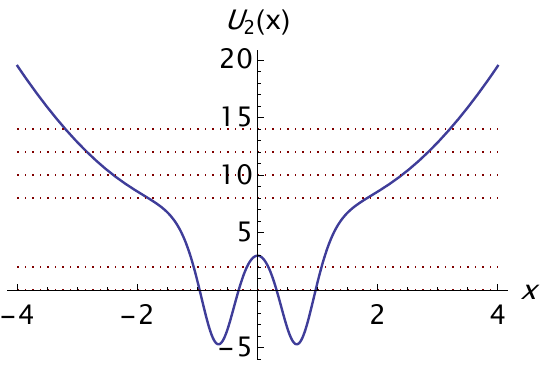}\quad

\includegraphics[scale=0.5]{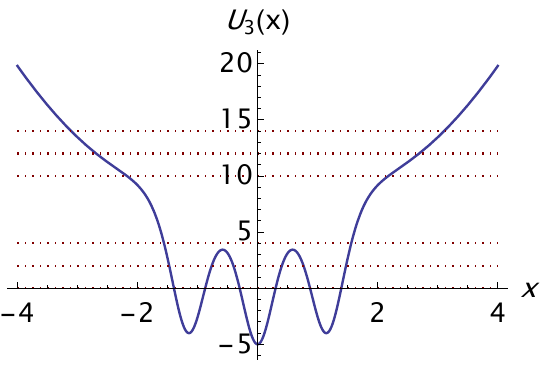}\quad
\includegraphics[scale=0.5]{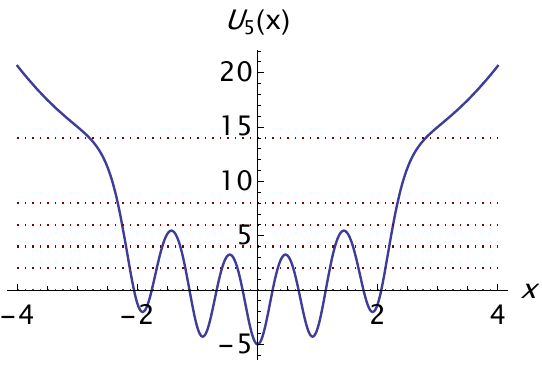}\quad
\end{center}
\caption{\small Plots of the deformed harmonic oscillator potential (\ref{UJ}) for different values of $J$. In the first line we have only the cases $J={2,4}$, while in the second line we have the cases
$J={3,5}$.}
\label{figU2}
\end{figure}

By using the Wronskian formula in (\ref{map}) we get the eigenstates of the system and its spectrum. Their explicit forms are given by
\begin{align}
&\psi_{n,J}(x)=N_{n,J}\frac{ R_{n,J}(x)}{P_J(x)}e^{-\frac{x^2}{2}}\;,&&
\omega_n^2=2n\;,&& n=0,1,\ldots J-1\,,\\
&\psi_{n,J}(x)=N_{n+2,J}\frac{ R_{n+2,J}(x)}{P_J(x)}e^{-\frac{x^2}{2}}\,,&&
\omega_n^2=2(n+2)\;,&&n=J,\ldots,\\
&N_{n,J}=\frac{1}{2 \sqrt[4]{\pi } \sqrt{J (J+1)} \sqrt{2^n n!}}\;,
\end{align}
and
\begin{equation}
\begin{aligned}
\nonumber
R_{n,J}(x)=&
(n-1) n P_J(x) H_{n-2}(x)+\\&
+J (J+1) H_{J-1}(x) (J H_{J-1}(x) H_n(x)-n H_J(x) H_{n-1}(x))+\\
&-(J-1) J H_{J-2}(x) ((J+1) H_J(x) H_n(x)-n H_{J+1}(x) H_{n-1}(x))\;.
\end{aligned}
\end{equation}
 The levels of this system are characterized by two towers of states separated by a gap $\delta\omega^2=\omega_{J}^2-\omega_{J-1}^2=4$. The tower below the gap contains $J$ states. In the Figure \ref{figU2} we show the first examples of such deformed Schr\"odinger potentials.

The Gaussian factor in the states confirms that they are bounded, so the situation is similar to the previous two cases. On the other hand, since the scalar potential and the stationary solutions are given in terms of the ground state of the Schr\"odinger system, it is worth taking a look at it closer,
\begin{align}
\label{psi00}
&\psi_{0,J}(x)=\frac{2 \sqrt{J (J+1)}}{\sqrt[4]{\pi }}\frac{ J H_{J-1}(x){}^2-(J-1) H_{J-2}(x) H_J(x)}
{ (J+1) H_J(x){}^2-J H_{J-1}(x) H_{J+1}(x)} e^{-\frac{x^2}{2}}\;,
\\
&\psi_{0,2k-1}(0)=\frac{ \sqrt{2k}}{\sqrt[4]{\pi } \sqrt{2 k-1}}\;,\qquad
\psi_{0,2k}(0)=\frac{ \sqrt{2k}}{\sqrt[4]{\pi } \sqrt{2 k+1}}\;,\qquad k=1,2,\ldots\;.
\end{align}
Here, for the sake of simplicity, we have split the values for index $J$ in even and odd cases. In the rest of the analysis, it will be apparent that the parity of this index fixes the nature of the scalar theories we will construct.

\begin{figure}[H]
\begin{center}
\includegraphics[scale=0.55]{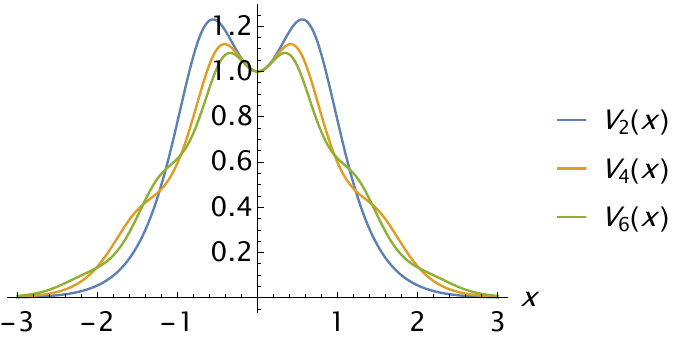}\quad
\includegraphics[scale=0.55]{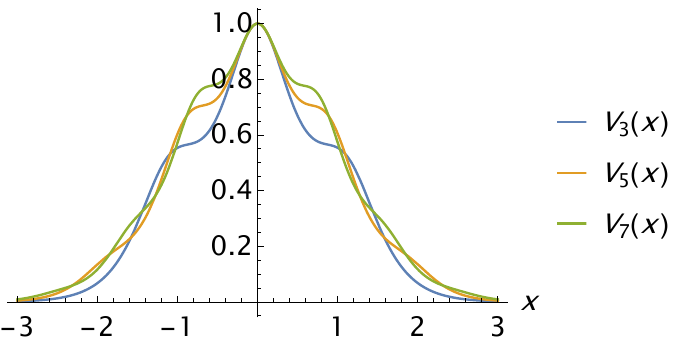}\quad
\end{center}
\caption{ \small Plots of the potential $V(x$), as a function of $x$. In the left panel, we plot the resulting potentials for odd $J$.
 In the right panel the form of the potentials for even $J$ are shown.}
\label{figV}
\end{figure}

Let us now derive some of the properties of the scalar theory associated with these types of Schr\"odinger systems. In the first step, using the relation (\ref{GenKink}), we can construct the potential \( V(x) \). The resulting forms for the initial cases of \( J \) are illustrated in Figure \ref{figV}.

\begin{figure}[H]
\begin{center}
\includegraphics[scale=0.5]{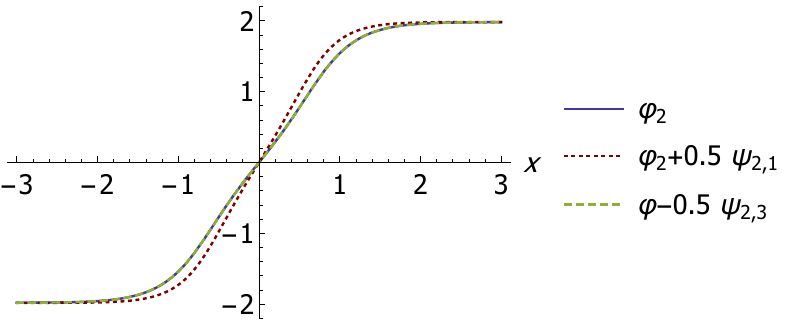}\quad
\includegraphics[scale=0.5]{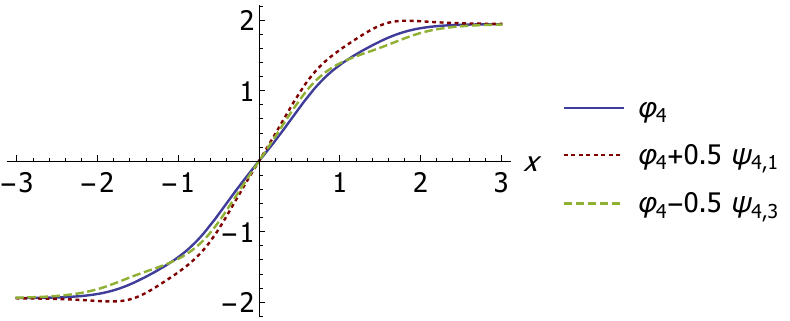}\quad
\\
\includegraphics[scale=0.5]{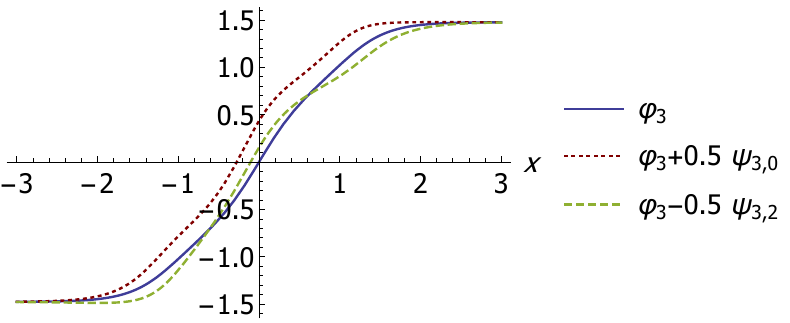}\quad
\includegraphics[scale=0.5]{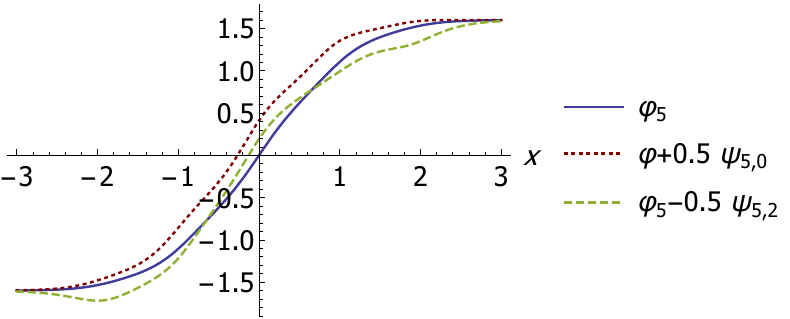}\quad
\end{center}
\caption{\small Plots of the stationary solutions of the form (\ref{GenKink}) for different values of $J$. In the first line we have only the cases $J={2,4}$, while in the second line we have the cases
$J={3,5}$.}
\label{figK}
\end{figure}

It is noteworthy that in the even cases  $J = 2k$, the center of the potential is a local minimum (different from zero), while two local maxima are present. As $k$ increases, these maxima diminish, and the potential takes on a more intricate, weaving shape. In the odd cases $J = 2k - 1$, instead of local maxima, some inflection points emerge. As $k$ grows, these inflection points approach the global minimum at $x = 0$. These ``obstacles" may influence the form of stationary solutions. The second equation in (\ref{GenKink}) is used to construct these solutions. The plots shown in Figure \ref{figK} illustrate how the confined kink solutions are distorted by the characteristics of the potential (depicted by the solid line).
The evaluation of our kink-like solution at  $x\rightarrow \infty$ gives us the position of the vacuum for the potentials. They are numerically evaluated in the following tables for even and odd values of index $J$.
\begin{eqnarray}
\begin{array}{|c|c|c|c|c|c|c|c|c|c|c|}
\hline
J & 1 & 3& 5 & 7 & 9 & 11 & 13 & 15 & 17 & 19 \\
\hline
 \varphi_v & 0.967 & 1.476 & 1.601 & 1.650 & 1.677 & 1.694 & 1.706 & 1.715 & 1.721 & 1.727 \\
\hline
\end{array}
\\
\begin{array}{|c|c|c|c|c|c|c|c|c|c|c|}
\hline
J & 2 & 4 & 6 & 8 & 10 & 12 & 14 & 16 & 18 & 20  \\ \hline
\varphi_v & 1.981 & 1.944 & 1.902 & 1.874 & 1.856 & 1.842 & 1.833 & 1.826 & 1.820 & 1.815 \\
\hline
\end{array}
\end{eqnarray}
For odd  (even) $J$, $\varphi_v$ asymptotically grows (diminish) up to $\sqrt{\pi}\approx 1.772$. So it seems that as $J$ increases, the theory tends to the error kink model from above or below in dependence of $J$ parity.

\begin{figure}[H]
\begin{center}
\includegraphics[scale=0.5]{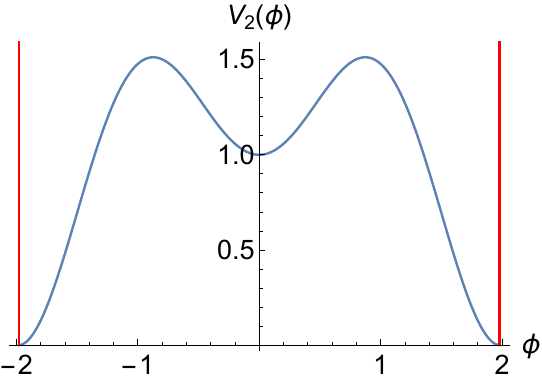}\quad
\includegraphics[scale=0.5]{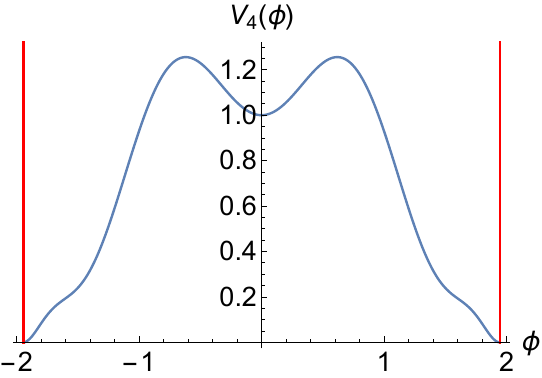}\quad
\\
\includegraphics[scale=0.5]{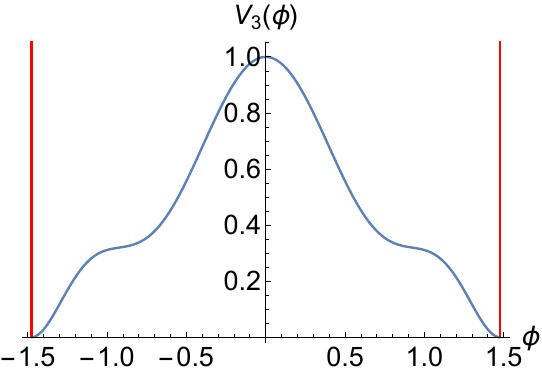}\quad
\includegraphics[scale=0.5]{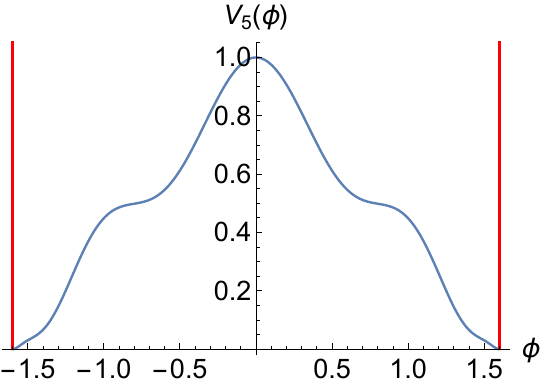}\quad
\end{center}
\caption{ \small Plots of the potential as a function of $\phi$ for some values of $J$. In the first line we have only the cases $J={2,4}$, while in the second line we have the cases
$J={3,5}$.}
\label{figVf}
\end{figure}

By compiling  all the information and using the inverse function of the deformed kinks, we numerically plot the potentials as a function of the field. These plots are displayed in the Figure \ref{figVf}.

As we have expected, the resulting potential $V(\phi)$ remains confining. However, the rational term in  $U(x)$ introduces an additional parameter $J$ that allows for fine-tuning the shape of the kink profile while preserving its structural stability and the purely discrete nature of its fluctuation spectrum.

To finish the construction at the classical level, we employ the last relation in (\ref{GenKink}) to compute the classical mass. After splitting again $J$ in even and odd values, we get
\begin{equation}
M_{2k-1}=M_0 \left(1-\frac{1}{2k}\right)\;,
\qquad
M_{2k}=M_0 \left(1+\frac{1}{2k}\right)\;,
\end{equation}
which clearly shows how the mass tends to the error kink mass $M_0$ when $k\rightarrow \infty$.

These results so far may imply that if the gap in the frequency spectrum appear far away the ground frequency $\omega_0=0$, observables such as the classical mass at rest and the position of the vacuum may look very similar to those of the error kink model, even though the corresponding scalar theories look very distorted in comparison with the error kink themselves. However, the differences must be relevant in quantum theory. This fact can be seen by looking at the corresponding spectral-$\zeta$ functions for the deformed models,
\begin{equation}
\zeta_{J,\omega}(s)=\frac{1}{2^{\frac{s}{2}}}\left(\sum_{n=1}^{J-1}\frac{1}{n^{\frac{s}{2}}}+\sum_{n=J+2}^{\infty}\frac{1}{n^{\frac{s}{2}}}\right)=\frac{1}{2^{\frac{s}{2}}}\left(\zeta(s/2)-\frac{1}{J^{\frac{s}{2}}}-\frac{1}{(J+1)^{\frac{s}{2}}}\right)\;,
\end{equation}
and by evaluating it at $s=-1$ we obtain the value of the quantum correction
\begin{equation}
Q= \frac{\hbar \lambda}{\sqrt{2}}\left(\zeta(-1/2)-\sqrt{J}-\sqrt{J+1}\right)\;.
\end{equation}
So, in contrast with what occurs for the classical observables, the quantum correction becomes more relevant as $J$ increases. Moreover, the factor can quickly dominate over the term related to the error kink correction.

The generalization of this picture for more complicated deformations is straightforward. The following generic election of the seed states
\begin{equation}
\Delta=\{\psi_{n_1},\psi_{n_1+1},  \psi_{n_2},\psi_{n_2+1},\ldots, \psi_{n_k}\psi_{n_k+1} \}\;,\qquad n_i\not=0
\end{equation}
produce a non-singular deformation of the rational harmonic oscillator with internal gaps, see \cite{Krein,Adler}. The missing energy level, as well as the missing degrees in the series of exceptional orthogonal polynomials corresponds to the sequence $n_{i}$ with $i=1,\ldots,k$. The ground energy state will have the form
\begin{equation}
\label{psigroundg}
\psi_0^{\Delta}\approx \frac{W(\psi_{n_1},\psi_{n_1+1}, \ldots, \psi_{n_k}\psi_{n_k+1},\psi_0)}{W(\psi_{n_1},\psi_{n_1+1}, \ldots, \psi_{n_k}\psi_{n_k+1})} \approx \frac{W(H_{n_1-1},H_{n_1},\ldots, H_{n_k-1},H_{n_k})}{W(H_{n_1+1},\psi_{n_1+2},  \ldots, H_{n_k}H_{n_k+1})}e^{-\frac{x^2}{2}}
\end{equation}
while other eigenstates are given by
\begin{equation}
\psi_\ell^{\Delta}\approx \frac{W(H_{n_1},H_{n_1+1}, \ldots, H_{n_k}H_{n_k+1},H_{\ell})}{W(H_{n_1-1},H_{n_1},\ldots, H_{n_k-1},H_{n_k})}\psi_0^{\Delta}\;,\qquad
\ell\not=n_i\;.
\end{equation}
All the features of the scalar final model we aim to construct are derived from the formulas (\ref{GenKink}). In this context, the local maxima and minima structure is determined by (\ref{psigroundg}), along with the deformations associated with the confining kink. All states are bounded and normalized for construction, and there is no zero-point mass renormalization. At the quantum level, the spectral-$\zeta$ function will be defined by
\begin{equation}
\zeta_\omega(s)=\frac{1}{2^{\frac{s}{2}}}\sum_{j}^\infty \frac{1}{j^\frac{s}{2}}\;,\qquad
j\in \N-\{n_1,n_2,\ldots,n_{k},n_{k+1}\}\;.
\end{equation}
We can rewrite this expression by adding and substantiating terms of the form  $\frac{1}{2^{\frac{s}{2}}}\frac{1}{n_i^s}$. This yields as a result
\begin{equation}
\zeta_\omega(s)=\frac{1}{2^{\frac{s}{2}}}\left(\sum_{j=1}^\infty \frac{1}{j^\frac{s}{2}}-\sum_{i=1}^{k+1} \frac{1}{n_j^\frac{s}{2}}\right)\;,\qquad
Q= \frac{\hbar \lambda}{\sqrt{2}}\left(\zeta(-1/2)-\sum_{i=1}^{k+1} \sqrt{n_j}\right)\;.
\end{equation}
Therefore, any deformation negatively impacts quantum correction, indicating that the quantum attractive force of the confining kink increases with each parity of subsequent perturbation modes eliminated within the Darboux transformation procedure.

\section{Discussion and Outlook}
\label{Sec_Discussion}

In this article, we reconsider the stability analysis of scalar theories to define a new exotic model and introduce solutions, which we called ``confining kinks''. These solitary waves are characterized by their perturbations, which are harmonic and bounded from below, meaning they possess a discrete spectrum of frequencies. After quantization, these models exhibit the absence of free meson propagation in their vacuum states, and zero-point renormalization does not apply. However, we can still compute finite one-loop quantum corrections to the mass at rest using the zeta function regularization scheme \cite{Zeta}, in a similar sense to techniques used to compute the Casimir effect  \cite{Bordag2001}.

The first model we construct is the error kink model. For this purpose, we identify the evolution equation for perturbations with the shifted quantum harmonic oscillator Schr\"odinger equation. It is well known that the ground state of this Schr\"odinger system is represented by a normalized Gaussian function. Consequently, the kink and its associated potential are expressed in terms of the error function and its inverse. We can analytically compute the mass and the expectation values of the vacuum. Additionally, the quantum correction to the mass is given in terms of the Riemann-$\zeta$ function evaluated at \( z = -\frac{1}{2} \).

Within the framework of resurgence theory, the divergence of $V''(\phi)$ at the minima indicates that perturbative fluctuations around the vacuum are ill-defined or non-summable in the Borel sense. Future investigations should determine whether, in the interpretation of the error kink at $N\rightarrow \infty$ of $V[\phi, N]$, the Borel singularities of the vacuum sector merge at $V(\phi)=0$, thereby confining the dynamics \cite{Dunne1999}.

To construct similar models, we consider the Darboux transformation. This technique acts on a Schr\"odinger Hamiltonian operator, producing a new quantum system along with its eigenstates. When applied to the quantum harmonic oscillator, it results in what are known as rational deformations. These operators exhibit gaps in their spectrum, and their solutions are expressed in terms of exceptional orthogonal polynomials \cite{quesne,quesne2,Gomez}. In this vein, we pick such rational deformation as the evolution operator of the corresponding perturbations. The scalar models derived from this approach introduce deformations into the potential and kink solutions based on the number of gaps in the frequency spectrum. Generally, the spectral-$\zeta$ function associated with these models, and consequently, the one-loop correction of the mass, is equal to the spectral-$\zeta$ function of the undeformed system minus the energy levels corresponding to the gaps in the spectrum (degrees of deformation). The simplest deformations are associated with the Owen kink model, named after the Owen function commonly used in statistics \cite{Owen}.

To continue this exploration, one can consider any Schr\"odinger system with a normalizable ground state. It is well-known that the P\"oschl-Teller reflectionless systems relate to the sine-Gordon and \(\Phi^4\) theories \cite{AMP}. In this context, it would be interesting to investigate the characteristics of a kink whose perturbations are governed by the Schr\"odinger operator with a Dirac delta function potential or by a system with an infinite potential well. In any case, such kinks may be topological defects associated with their respective topological charges
\begin{equation}
N=\varphi(\infty)-\varphi(-\infty)=2\varphi_v\;,
\end{equation}
see \cite{Vile}, so we may think on applications of them in
condensed matter physics, non-linear optics or in cosmology. First, by following the approach outlined in \cite{semenoff},  one-dimensional domain wall based on our kinks can be induced in a Dirac topological insulator. In the continuous approximation, the corresponding tight-binding model Hamiltonian yields
$
H_D=-\Ti \hbar v_f(\sigma_1\partial_x+\sigma_2\partial_y)+m\varphi(x)\sigma_3\,.
$
Here, $m$ is a constant with mass dimension, $\sigma_i$ denote the Pauli matrices, and $v_f$ represents the Fermi velocity. Such a material exhibits topologically protected zero-energy bound states, as demonstrated by the index theorem and the associated vacua of the kink \cite{Horsley_2018}. Recent advances in technology enable the atom-by-atom construction of these devices using various experimental techniques \cite{chang2018colloquium}. Investigating the effects of total confinement at this scale, particularly in comparison with $\Phi^4$ theories, remains an open area of research. In optics, spectral (supersymmetric) design is already employed to create distinct light paths that mimic quantum mechanical behavior  by identification the refraction index with the potential \cite{Garbin_2021,huang2022guiding}. Moreover,  soliton-type index barriers have been explored in \cite{Miri_2013}. In cosmology, theoretical predictions suggest that domain walls may form in the early universe \cite{Vile}. The presented models facilitate the study of non-radiating domain walls through perturbative computations in lattice numerical simulations.

Along the same lines, generalizing confining kinks to higher dimensions presents an interesting prospect. Although Derrick's theorem precludes stable topological defects in purely scalar potentials in higher dimensions \cite{Rubakov}, this obstacle can be bypassed by introducing couplings to fermions or gauge fields. Two alternative approaches are worth noting: the first extends the Jackiw-Rebbi model \cite{Jackiw} which find applications in the study of polymeric chains \cite{JS}. In particular it would be interesting to investigate the effects to change the $\Phi^4$ potential in this kind of models by confining kink potentials.  The second approcuh adopts a supersymmetric path \cite{witten} utilizing the superpotential $W(\varphi)$ from (\ref{MW}). The Lagrangian should be of the form
\begin{eqnarray}
{\cal L}&=&\frac{1}{2}\partial_\mu \phi \partial^\mu \phi-\frac{1}{2}\frac{\partial W}{\partial \phi}\cdot \frac{\partial W}{\partial \phi}+\Ti \overline{\Psi}\gamma^\mu \partial_\mu \Psi -\frac{1}{2}\overline{ \Psi}\frac{\partial^2  W}{\partial \phi^2} \Psi \;,
\end{eqnarray}
where $ \Psi=\Psi(x,t)$ is a Majorana spinor and   $ \gamma^\mu$ are the Dirac matrices.
The stability analysis for this case remains the same for the scalar field, while the fermions perturbations are now governed by Dirac operators \cite{izquierdo2004quantum}. Such operator possess a natural supersymetric structure, and it also can be modified by its proper version of the Darboux transformation \cite{Samsonov,Samsonov2}. We will carry with this investigation in further publications.

We emphasize the dynamical implications of our inverse construction regarding the complete absence of a continuous spectrum. In standard topological soliton frameworks, time-dependent perturbations or scattering processes typically cause the kink to dissipate energy through the radiation of mesonic modes into the continuum. In our models, this conventional dissipation mechanism is inherently suppressed due to the purely discrete nature of the linear fluctuations. Consequently, the newly found configurations exhibit a remarkable ``structural rigidity'' against radiative decay. While this structural feature naturally restricts the study of conventional asymptotic scattering dynamics, it opens an alternative avenue for investigating topological defects characterized by exceptional stability against thermal or radiative fluctuations. This property could find valuable applications in both condensed-matter analogues and early-universe cosmological scenarios.

Additionally, it is worth exploring the quantization of such a confining system from the Feynman path integral approach point of view to obtain information about thermodynamic quantities. Since the field is now restricted in value by the vacuum point (considering the infinite walls beyond these points), some restrictions should appear in the definition of the measure of integration. Therefore, studying the path integral approach of these confining systems would require extending the conventional formalism by incorporating some mathematical tools such as \cite{Asorey2007,Merdaci2014}. Another possibility is to induce some restriction in the measure of the integral, something reminiscent of the Gribov horizon (see \cite{Gribov1978,Sobreiro2005,Dudal2008}). As our technique give us a confined field almost by construction, the restriction on the integration range in the functional space, besides the connection to the restriction in the integration limits, also could give us some light on the confinement phenomena. We plan to come back to these intriguing questions in a forthcoming work.

\subsection*{Acknowledgements}
 L.~I. was supported by Fondecyt Grants No. 3220327. L.~I. would like to thank Prof. Francisco Correa for his insightful  comments. L.~I. also thanks the hospitality of the Universidad de Salamanca, the Universidad de Valladolid, and the Universidad Austral de Chile, where this research began. P.~P. gladly acknowledge support from Charles University Research Center (UNCE 24/SCI/016).

\printbibliography[title={References}]


 \end{document}